\documentstyle[aps,prl,twocolumn]{revtex}
\input epsf

\begin{document}
\draft
\title{Vortex nucleation in Bose-Einstein condensates in an oblate, purely magnetic potential}
\author{E. Hodby, G. Hechenblaikner, S.A. Hopkins, O.M. Marag\`o and C.J. Foot}
\address{Clarendon Laboratory, Department of Physics, University of Oxford,\\
Parks Road, Oxford, OX1 3PU, United Kingdom.}
\date{\today}

\maketitle

\begin{abstract}
We have investigated the formation of vortices by rotating the
purely magnetic potential confining a Bose-Einstein condensate. We
modified the bias field of an axially symmetric TOP trap to create
an elliptical potential that rotates in the radial plane. This
enabled us to study the conditions for vortex nucleation over a
wide range of eccentricities and rotation rates.
\end{abstract}


The existence of quantised vortices is one of the most striking
and fascinating signatures of superfluidity. There has been
theoretical interest in vortices in a dilute gas BEC for many
years (for a review see for example \cite{fetter}) and the recent
experimental observations \cite{Jila1,Paris1} have stimulated much
further work. Vortices have been nucleated using phase imprinting
\cite{Jila1} and by using a laser beam
'stirrer'\cite{Paris1,MIT1}. We have used a new, purely magnetic
excitation scheme, which like the optical stirrer has a close
analogy to the 'rotating bucket' experiment with liquid He
\cite{He}. Vortices provide dramatic visual evidence for the
single condensate wavefunction, the detailed mechanisms for vortex
formation, stabilisation and decay reveal new information about
the dynamics of the excited states of the condensate and its
interaction with the thermal cloud. The excited states and hence
the conditions for nucleation depend on the geometry of the
confining potential \cite{Stringari}. Our work, in an oblate
geometry compliments the work done to date in prolate and
spherical traps. The range of eccentricity available in our purely
magnetic rotating trap \cite{Arlt} is much greater than in
previous experiments, where the radial symmetry was broken by
`stirring' the condensate with the dipole force from off-resonant
laser beams.

We have measured the relationship between the eccentricity of the
trap in the plane of rotation and the angular frequency $\Omega$
that must be satisfied to nucleate vortices in our trap. The trap
has a harmonic potential with oscillator frequencies $\omega_x <
\omega_y < \omega_z$. The shape of the trap is characterised by
the deformation parameter $\epsilon$, given by
\begin{center}
\begin{equation}
\epsilon = \frac{\omega_y^2 - \omega_x^2}{\omega_x^2 + \omega_y^2}
\end{equation}
\end{center}
We show that when the trap is adiabatically ramped from circular,
$\epsilon=0$, to a given eccentricity, there is both an upper and
lower critical rotation rate for nucleation. We present the
threshold conditions for vortex nucleation in our geometry and
finally discuss the role of the thermal cloud in both the
nucleation and stabilization processes.

A detailed description of the rotating elliptical trap is given in
\cite{Arlt} and only a summary of the technique is presented here.
Our TOP trap consists of a spherical quadrupole field and a
rapidly rotating bias field \cite{TOP}. The zero of the magnetic
field describes a circle in the radial (XY) plane. The atoms
experience a time-averaged, axially symmetric harmonic potential.
If the X and Y components of the bias field ($B_x$ and $B_y$) are
not of equal amplitude, then the axial symmetry is broken and the
potential becomes elliptical in the XY plane. The ratio of the
radial trap frequencies $\omega_x / \omega_y $, can be calculated
numerically from $B_y / B_x$. For small eccentricities $\omega_x /
\omega_y -1 \approx (B_y / B_x - 1)/4$. To rotate the elliptical
potential we modulated $B_x$ and $B_y$ at $\Omega$, a frequency
much lower than the TOP frequency $\omega_0$. The final bias field
has the form of a rotation matrix through angle $\Omega t$
multiplying the X and Y components of an ellipse:

\begin{center}
\begin{equation}
\left(
\begin{array}{c}
B_x \\ B_y
\end{array}
\right) = \left(
\begin{array}{cc}
\cos \Omega t & \sin \Omega t \\ -\sin \Omega t & \cos \Omega t
\end{array}
\right) \left(
\begin{array}{c}
E B_t \cos\omega_0 t \\ B_t \sin\omega_0 t
\end{array}
\right)
\end{equation}
\end{center}
During any particular evaporative cooling run the value of
$\Omega$, the angular frequency at which the elliptical trap
rotates, was fixed (for technical reasons) and so the trap was
always spinning. To create a 'static' trap for efficient
evaporative cooling, the bias field ratio, E, is set to 1 so that
the trap is symmetric around the rotation axis. This symmetry was
checked to be accurate to $\epsilon = 0 \pm 0.005$ using
measurements of the frequency of dipole oscillations in the X and
Y directions. Evaporative cooling followed by an adiabatic
expansion gave a condensate of $ 2 \times 10^4$ atoms, at a
temperature of $0.5 T/T_c$. At this stage the trap is axially
symmetric, $\omega_x =\omega_y$, with trap frequencies typically
$\omega_x/2\pi = 62$~Hz and $\omega_z/2\pi = 175$~Hz.

\begin{figure}
\begin{center} \mbox{ \epsfxsize 3in\epsfbox{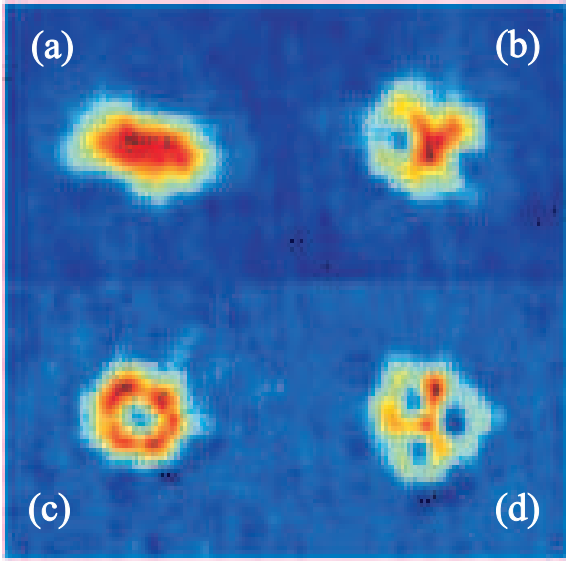}}
\end{center}
\caption{Images of the condensate at different stages during the
vortex nucleation process. (All taken with $\overline{\Omega} =
0.70$ and $\epsilon$ = 0.05 and after 12ms of free expansion). (a)
After the 200ms spinning eccentricity ramp, the condensate is
elongated, indicating that a quadrupole mode has been excited. (b)
After a further 600ms in the spinning trap one vortex has just
formed near the edge. After 800ms, the vortices have reached their
equilibrium positions and appear in symmetrical configurations.
(c) shows one centred vortex, typical under these conditions,
whilst (d) shows a triangular array of three
vortices.}\label{vortall2}
\end{figure}

To create vortices, the value of E was ramped linearly over 200ms
from 1 to its final value, to give a trap that was elliptical in
the rotating frame. If one creates an eccentric trap by increasing
(decreasing) the TOP bias field then all three trap frequencies
are reduced (increased). Since vortex lifetimes depend on the mean
trap frequency \cite{fedichev}, we adjusted the quadrupole field
in the adiabatic expansion stage, to ensure that all traps had the
same average radial trap frequency $\omega_{\perp}$, defined as
\begin{center}
\begin{equation}
\omega_{\perp} = \sqrt{\frac{\omega_x^2 + \omega_y^2}{2}}.
\label{perp}
\end{equation}
\end{center}
The condensate was then held in the rotating anisotropic trap
\cite{Arlt} for a further 800~ms before being released. After 12ms
of free expansion the cloud was imaged along the axis of rotation
using an absorption imaging system with 3~$\mu$m resolution.
Figure \ref{vortall2} shows images of the expanded condensate at
different stages during the nucleation process. Initially the
cloud elongates, confirming that nucleation is being mediated by
excitation of a quadrupole mode (fig.\ref{vortall2}(a))
\cite{Paris4}. Then finger-like structures appear on the outside
edge of the condensate which eventually close round and produce
vortices, $\sim 800$ms after rotation began
(fig.~\ref{vortall2}(b)). Approximately 200~ms later, these have
moved to equilibrium positions within the bulk of the condensate.
Figures \ref{vortall2}(c) and (d) show typical, single-shot images
of stable vortex arrangements. The depth of each vortex (in the
integrated absorption profile) is approximately $95 \%$ of the
surrounding condensate. The core diameters of $\sim 5 \mu$m after
12~ms of free expansion are consistent with the predictions in
\cite{Lundh}. The maximum number of vortices we observed was
seven, limited by the number of atoms in our condensate.

\begin{figure}
\begin{center}\mbox{ \epsfxsize 3in\epsfbox{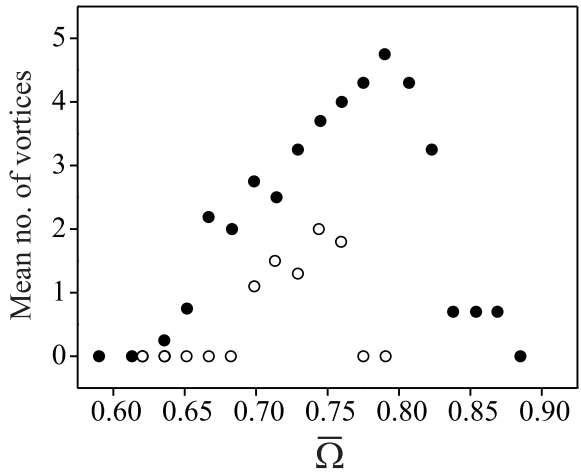}}
\end{center}
\caption{The mean number of vortices as a function of the
normalised trap rotation rate $\overline{\Omega} =
\Omega/\omega_{\perp}$. Two different trap eccentricities were
used, $\epsilon$=0.041 (open circles) and $\epsilon$=0.084 (solid
circles). Each data point is the mean of 4 runs.
}\label{scanomega2}
\end{figure}

Our first study of the nucleation process involved counting the
number of vortices as a function of the normalised trap rotation
rate, $\overline{\Omega} = \Omega/\omega_{\perp}$, for a fixed
eccentricity. Results for trap deformations $\epsilon$ = 0.084 and
0.041 are given in fig.~\ref{scanomega2}. These graphs show a
maximum and minimum value of $\overline{\Omega}$ for nucleation at
a given eccentricity. Increasing $\epsilon$ increases the range of
$\overline{\Omega}$ over which vortices may be nucleated, both by
lowering $\overline{\Omega}_{min}$ and raising
$\overline{\Omega}_{max}$. In the limit of small eccentricities,
$\sqrt{2}\omega_{\perp}$ is the frequency of the m=2 quadrupole
mode, which has been shown elsewhere to play a critical role in
the nucleation process \cite{Paris4}. Rotation of the ellipsoidal
trap at half this frequency, i.e.
$\overline{\Omega}_c=1/\sqrt{2}\simeq 0.71$, excites this mode.
The plots in fig.~\ref{scanomega2} confirm that the nucleation
depends on resonant excitation of the quadrupole mode as seen
previously \cite{MIT1,Paris4}. The resonance is broader at higher
eccentricity, as intuitively expected for stronger driving.

\begin{figure}
\begin{center}\mbox{ \epsfxsize 3in\epsfbox{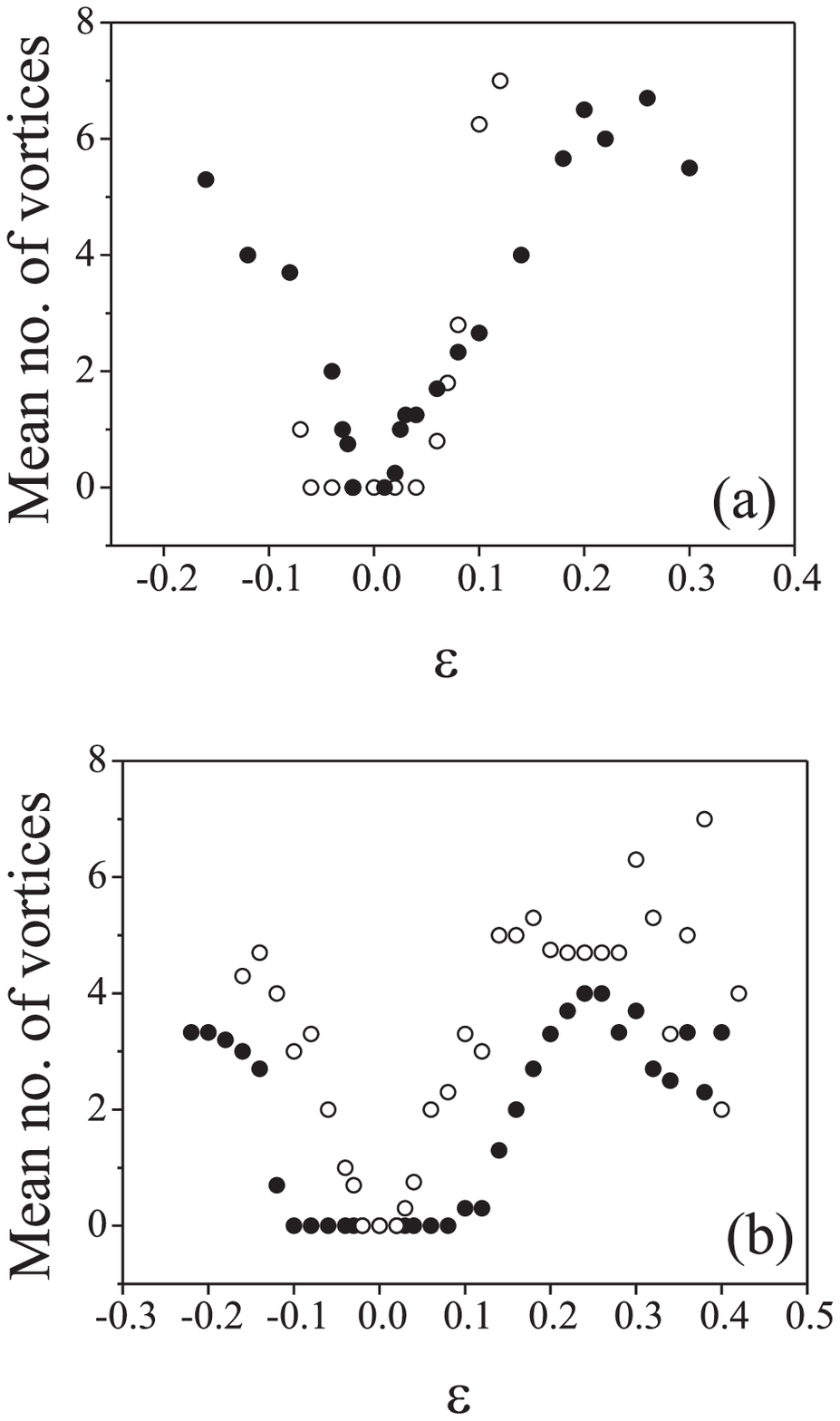}}
\end{center}
\caption{The mean number of vortices as a function of trap
deformation at 4 different trap rotation rates: (a) above and (b)
below the critical value $\overline{\Omega}_c=0.71$. In (a),
$\overline{\Omega}= 0.74$ (solid circles)and $0.81$ (open
circles). In (b) $\overline{\Omega}= 0.61$ (solid circles) and
$0.70$ (open circles). Positive (negative) $\epsilon$ corresponds
to $\omega_x < \omega_y$ ($\omega_x > \omega_y$).}\label{scanecc}
\end{figure}

Our second study involved holding $\overline{\Omega}$ constant and
counting the number of vortices as a function of the trap
deformation $\epsilon$. Figure \ref{scanecc} shows our results for
two cases: (a) $\overline{\Omega}>\overline{\Omega}_c$ and (b)
$\overline{\Omega}<\overline{\Omega}_c$. Interestingly we were
able to nucleate vortices under adiabatic conditions when
$\overline{\Omega} < \overline{\Omega}_c$. We employ the
adiabaticity criterion $\dot{\epsilon}/\epsilon \ll
\omega_{\perp}$. As a further check of the adiabaticity we varied
the time for the eccentricity ramp between $200$~ms and $1$~s, and
detected no difference in the number of vortices formed.

\begin{figure}
\begin{center}\mbox{ \epsfxsize 3in\epsfbox{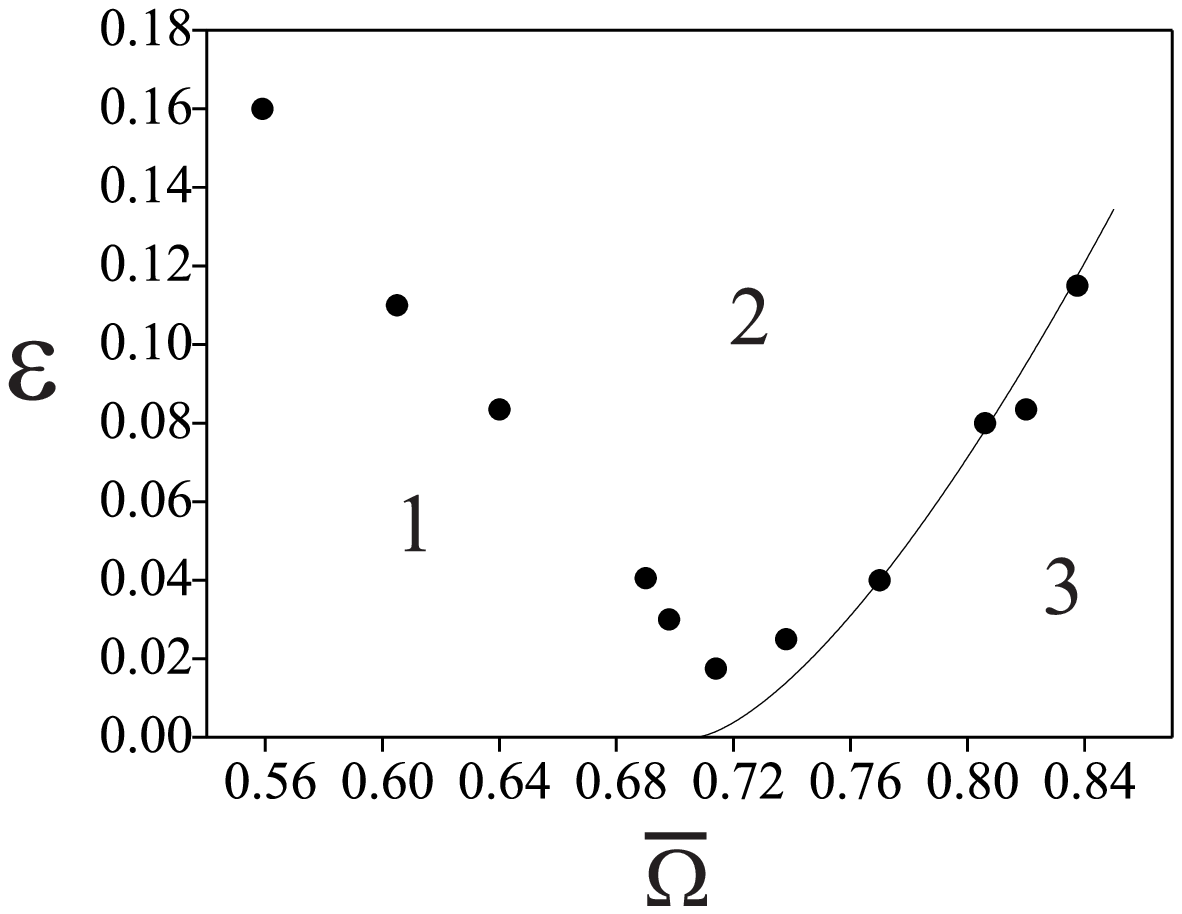}}
\end{center}
\caption{The critical conditions for vortex nucleation. The data
points mark the minimum trap deformation for nucleation at a
particular $\overline{\Omega}$. Vortices may be formed in region
2. The solid line shows the theoretical limit of stability of the
quadrupole mode, which is stable in region 3. This line is in good
agreement with the extreme conditions for vortex formation at
$\overline{\Omega}
> \overline{\Omega}_c $}  \label{nucleate2}
\end{figure}

The critical values of $\overline{\Omega}$ and $\epsilon$ for
nucleation were extracted from plots such as fig.~\ref{scanomega2}
and fig.~\ref{scanecc} for many values of $\overline{\Omega}$ and
compiled on fig.~\ref{nucleate2}. The data points show the minimum
eccentricity required for a given rotation rate and map out region
2, within which vortices nucleate. $\overline{\Omega}_c$ appears
to be a critical rotation frequency at which vortices can be
nucleated with minimum eccentricity, as predicted in
\cite{Castin}. Changing $\overline{\Omega}$ in either direction
requires a more elliptical trap for nucleation, although different
physical processes control the upper and lower limits as explained
below.

At $\overline{\Omega} > \overline{\Omega}_c$, the condensate
follows a particular quadrupole mode at small eccentricity, region
3 in fig.~\ref{nucleate2}. This mode is referred to as the
'overcritical branch' in \cite{Stringari2} and has an elliptical
density distribution which is orthogonal to the trap potential. It
then nucleates vortices when the eccentricity is too large for the
quadrupole mode to be a solution of the hydrodynamic equations.
The boundary of the region in the $\epsilon$ versus
$\overline{\Omega}$ plot where this quadrupole mode exists is
given by

\begin{center}
\begin{equation}
\epsilon = \frac{2}{\overline{\Omega}} \left( \frac{2
\overline{\Omega}^2 - 1}{3} \right)^{3/2}.
\end{equation}
\end{center}
We determined this relation from the solutions of the hydrodynamic
equations for superfluids and it is plotted as a solid line in
fig.~\ref{nucleate2}. This line agrees well with the experimental
data for the critical conditions for nucleation for
$\overline{\Omega} > \overline{\Omega}_c$ and a wide range of
$\epsilon$.

Below $\overline{\Omega}_c$, the deformation needed to nucleate
vortices appears to increase linearly with $\overline{\Omega}$.
This boundary cannot be explained in terms of the stability limit
of a quadrupole mode - the 'normal branch' is stable on both
regions 1 and 2 and the 'overcritical branch' is stable in
neither. Our data appears to be at variance with the results in
\cite{Paris4}, where no vortices were seen when the eccentricity
was increased adiabatically and $\overline{\Omega} <
\overline{\Omega}_c$.

A mechanism for the creation of vortices at a frequency below
$\overline{\Omega}_c$ has been proposed in \cite{Castin}. They
have shown that there are regions in the plot of $\epsilon$ versus
$\overline{\Omega}$, both above and below $\overline{\Omega}_c$,
where the quadrupole solutions of the hydrodynamic equations
become dynamically unstable. For frequencies above
$\overline{\Omega}_c$ the predicted instability domains coincide
with the experimentally observed vortex domains in \cite{Paris4},
and this work, thus indicating a link between their instability
analysis and vortices. To make a quantitative prediction for the
boundary between regions 1 and 2 shown in fig.~\ref{nucleate2},
will require further detailed work for our specific case. We also
note that it is possible that there exist non-negligible terms of
higher order than quadratic in the rotating magnetic potential.
These have been demonstrated to excite high order rotating surface
modes, e.g. hexapole, which lead to vortex nucleation at
frequencies less than $\overline{\Omega}_c$ \cite{Paris4,MIT1}.

Another possible mechanism for observation of vortices below
$\overline{\Omega}_c$ is that the thermal cloud plays an important
role. Transfer of angular momentum to the condensate from the
spinning thermal cloud may provide a mechanism for vortices to
form at $\overline{\Omega} < \overline{\Omega}_c$. However the
transfer rate of angular momentum must be greater than any loss
rate due to residual trap anisotropy \cite{DGO}. In \cite{Paris4},
gravity produces a small static eccentricity in the trap in the
plane of rotation. The `spin down' time for a rotating thermal
cloud in the presence of a deformation parameter $\epsilon$ of
only $0.01$ is very short, 0.5~s, compared to the spin up time of
15~s and hence the cloud may never gain significant angular
momentum. However, in our experiment, gravity acts along the
rotation axis and hence the trap is symmetric in the plane of
rotation, giving a more favourable ratio of spin-up to spin-down
times.

With this hypothesis in mind, we tested our nucleation curve at
lower temperature to see if there was any change when the amount
of thermal cloud was reduced. When acquiring the data of
fig.~\ref{nucleate2} the rf was turned off after condensation and
some heating was observed during the nucleation procedure,
resulting in a temperature around $0.8 T_c$. To achieve a lower
temperature we left on the so-called `rf shield' so as to give an
effective trap depth of $800$~nK during the nucleation process.
This resulted in a temperature of $0.5 T_c$. No significant change
was observed in the nucleation curve at this lower temperature.
However, this does not totally rule out a role for the thermal
cloud in the nucleation process since even at $0.5 T_c$, there was
still a significant $20\%$ of atoms in the thermal cloud.

Although the exact amount of thermal cloud seemed to have little
effect on the nucleation conditions, it had a striking effect on
the behaviour of vortices after nucleation. Without the rf shield
during the nucleation process, vortices were only occasionally
found in an equilibrium configuration (i.e. 1 vortex in the
centre, 3 vortices in an equilateral triangle as in
fig.~\ref{vortall2}) and $\sim 400$ms after forming they had
already moved to the edge of the condensate before disappearing.
However with the rf shield present, the vortices were normally
found in equilibrium positions $\sim 200$~ms after formation and
had a lifetime of $\sim 4s$, only limited by the decay of the
condensate itself.

In summary, we have used a purely magnetic rotating trap to
investigate conditions for vortex nucleation (after the formation
of the condensate) over a wide range of trap eccentricities. For a
given eccentricity, we observe both an upper and lower limit to
the rotation rate for nucleation. The upper limit confirms the
predictions in \cite{Paris4}, but over a much wider range of
parameters. However the lower limit to the rotation rate is
different to that reported elsewhere. Further theoretical work is
required to explain the linear dependence of $\epsilon$ on
$\overline{\Omega}$ shown in fig.~\ref{nucleate2}. The thermal
cloud is shown to destabilise vortex arrays, thus measurements of
the vortex lifetime in an oblate geometry were made with an rf
shield to prevent heating.

The rotating anisotropic magnetic trap is well suited to
experiments that require a large trap eccentricity. It has
recently been used to investigate the irrotational behaviour of
vortex-free condensates at low rotation rates \cite{hech}. A large
eccentricity is also needed to spin up a thermal cloud \cite{DGO},
hence this trap may be useful in attempts to condense directly
into a vortex state from a spinning thermal cloud in thermodynamic
equilibrium.

We would like to thank D. Feder for useful discussions. This work
was supported by the EPSRC and the TMR program (No. ERB
FMRX-CT96-0002). O.M. Marag\`{o} acknowledges the support of a
Marie Curie Fellowship, TMR program (No. ERB FMBI-CT98-3077).

\end{document}